\documentclass[aps,prl,twocolumn,showpacs,floatfix]{revtex4}
\usepackage{epsfig}
\usepackage{graphicx}
\usepackage{dcolumn}
\usepackage{longtable}
\usepackage{amsthm,amsmath}

\begin{document}

\preprint{\today}

\title{Electric dipole moment of $^{225}$Ra due to P- and T-violating weak interactions} 

\author{Yashpal Singh{\footnote{yashpal@prl.res.in}} and B. K. Sahoo{\footnote{bijaya@prl.res.in}} }

\affiliation{Theoretical Physics Division, Physical Research Laboratory, Navrangpura, Ahmedabad - 380009, India}

\begin{abstract}
 We report rigorous calculations of electric dipole moment (EDM) in $^{225}$Ra due to parity and time-reversal 
violating tensor-pseudotensor (T-PT) and nuclear Schiff moment (NSM) interactions between the electrons and nucleus
by employing the relativistic all order coupled-cluster (RCC) methods at various levels of approximation. The most 
accurate EDM ($d_A$) results are obtained as $d_A=-10.04\times 10^{-20} C_T <\sigma_n>|e|cm$ and $d_A^{NSM}=-6.79 \times 
10^{-17} S (|e|fm^3)^{-1} |e|cm$ with $C_T$ and $S$ are the T-PT coupling constant and NSM respectively. Due to 
exhaustive treatment of the electron correlation effects in these calculations, the EDM results for the corresponding 
T-PT and NSM interactions reduce by about 45\% and 23\%, respectively, from the previously known values. 
Nonetheless they are still found to be 2-3 times larger than the $^{199}$Hg results. Validity of the RCC results 
are countenanced by comparing our calculations at the zeroth order Dirac-Fock method and all order random-phase 
approximation with the other groups. In view of the current experimental progress in the $^{225}$Ra EDM 
measurement, our high accuracy calculations are very useful to yield limits on $C_T$ and $S$ in this atom. Thusly,
it is possible to realize more accurate limits on the electron-quark T-PT interaction and $\theta_{QCD}$ parameter of
particle physics in future.
\end{abstract}


\maketitle

The experimental group at Argonne National Laboratory (ANL) has just communicated the first measurement of permanent 
electric dipole moment (EDM) of $^{225}$Ra \cite{lu} after delineating its progress gradually in the last several years
\cite{guest,scielzo,parker1}. They obtain an upper limit on the atomic EDM ($d_A$) as $|d_A$($^{225}$Ra)$|< 5.0
\times 10^{-22} |e|\text{cm}$ (at 95\% confidence). Though this limit is not competitive with the present current best 
limit available from $^{199}$Hg \cite{griffith} as $|d_A$($^{199}$Hg)$|< 3.1\times 10^{-29} |e|\text{cm}$ (at the same 
95\% confidence level), the ANL group portray their credence to surpass the limit on $d_A$($^{225}$Ra) over 
$d_A$($^{199}$Hg) sooner by making rapid improvement in their measurement \cite{holt}. Indeed, viability of such claims 
are conceivable from both the theoretical and experimental prospectives. From the theoretical notion, the octupole 
deformation can enhance EDM observation in $^{225}$Ra by a factor larger than $10^3$ compared to $^{199}$Hg \cite{auerbach,spevak}. 
Other theoretical advantages that favor to carry out $^{225}$Ra EDM measurement are, it has larger nuclear charge 
($Z$) compared to $^{199}$Hg and uncertainty due to the contribution from the octupole moment is still absent owing to 
its nuclear spin $I=1/2$. On the experimental scope, cold-atom techniques to measure Larmour precision have been
developed for $^{225}$Ra atoms \cite{guest,parker} that are least sensitive to the systematics \cite{romalis}. 
Moreover, the ANL research group also asserts to bring about significant improvement in the statistical uncertainty by 
increasing the number of atoms to $10^6$ by producing them using the facility for rare isotope beam (FRIB) with the
measurement time of 100 days \cite{holt}. 

 The EDM of a diamagnetic atom like $^{225}$Ra is sensitive to the parity (P) and time reversal (T) violating (P,T-odd) 
electron-nucleon tensor-pseudotensor (T-PT) and nuclear Schiff moment (NSM) interactions. The NSM originates 
primarily due to the distorted charge distribution inside the nucleus caused by the P,T-odd interactions among the 
nucleons mediated by the neutral pion ($\pi^0$-meson). The other possible source to the $^{225}$Ra EDM comes from the 
possible EDM or chromo-EDM of either up ($\tilde{d}_u$) or down ($\tilde{d}_d$) quark through the self-interactions in the $\pi$-meson loops \cite{pospelov,pospelov1}. Thus at the first hand if EDM of an atom
is being observed, it will be a clean signature of P,T-odd interaction inside the atom  \cite{khriplovich,landau}. 
The second but the most vital consequence is, it will paramount to observing semi-leptonic CP-violation via the 
CPT theorem \cite{luders} and can explain certain new physics beyond the standard model (SM) of particle physics. 
In the SM, the complex $\delta$-phase appearing in the Cabibbo-Kobayashi-Maskawa (CKM) mechanism is responsible for the 
hadronic CP-violation. Till date the CP-violation only in the hadronic sector has been observed \cite{christenson, alvarez}. The SM also predicts EDMs of elementary particles, but with sufficiently smaller 
values. However, it is largely believed that SM is an intermediate manifestation of a complete theory. Such insistences 
are strongly endorsed by observation of finite masses of neutrinos, matter-antimatter asymmetry in the Universe, 
existence of dark matter (indirectly) etc. \cite{dine,canetti}. Many models extension to SM have been propounded as 
attempts to perceive explanation to these unsettled inquests. Some of the prominent models capable of explaining
physics of the above problems are the left-right symmetric, supersymmetric (SUSY), multi-Higgs etc. models 
\cite{pospelov,engel,barr1}. These extensions to SM try to offer additional interactions, new particles and also incorporate 
other possible sources of CP-violation that can enhance EDMs of the elementary particles as well as of the composite
systems up to the level as par with the current sensitivity achieved by the atomic experiments. Thereby even if the 
limits of the T-PT coupling coefficients or NSM are improved from the atomic EDM studies, it can constrain to certain 
beyond SM models. In fact, NSM is related to $\theta_{QCD}$ parameter of the particle physics \cite{pospelov,yashpal-hg}
and improved value on $\theta_{QCD}$ can annex to new sources of CP-violation apart from the $\delta$-phase of SM which 
may be be sufficient enough to uphold the validity of those models that are able to justify simultaneously large value 
of $\theta_{QCD}$ and matter-antimatter asymmetry of the Universe. Therefore the ongoing efforts to improve limits on 
the extracted parameters from the atomic EDMs, even with the null results, are still of pragmatic use.

 With the current development in the EDM measurement of $^{225}$Ra, it is now absolutely necessary to perform the 
required calculations accurately so that better limits on the above discussed parameters can be deduced. Recently, we 
have developed and employed a number of many-body methods in the relativistic framework to calculate these quantities 
more reliably among which the relativistic coupled-cluster (RCC) methods are more powerful  \cite{yashpal-xe,bijaya-rn,
yashpal-hg}. In this paper, we present very accurate $d_A$ calculations for $^{225}$Ra due to both the T-PT ($d_A^{TPT}$) 
and NSM ($d_A^{NSM}$) interactions. We observe unusually large correlation effects in the determination of these properties, but exhibiting 
strong cancellations for which the final results are reduced substantially compared to the values reported previously
using the methods having facile approximations. In order to demonstrate the same, we also compare our results at the 
intermediate calculations considering the Dirac-Fock (DF) method and random-phase approximation (RPA). We, then, appraise
the role of the physical effects (specifically the non-RPA correlations) responsible for reducing the results that 
are usually appearing through the higher order perturbation theory. Moreover, we also perform calculation of the dipole 
polarizability ($\alpha_d$) of $^{225}$Ra by taking the electric dipole (E1) operator, which has identical angular 
momentum and parity selection rules with the T-PT and NSM interaction Hamiltonians, to compare its value with those of 
the previous calculations. This can provide additional insights into the potentials of various methods to yield accurate 
results.

The P,T-odd electron-nucleus (e-N) interaction Hamiltonians due to T-PT coupling and NSM are given by \cite{martensson,flambaum}
 \begin{eqnarray}
  && H_{e-N}^{T-PT}=\frac{iG_FC_T}{\sqrt{2}} \sum \mbox{\boldmath ${\vec \sigma}_n \cdot {\vec \gamma}_D$} \rho_n(r) \\
 \text{and} && \nonumber \\ 
  && H_{e-N}^{NSM}=\frac{3 \mbox{\boldmath ${\vec S} \cdot {\vec r} $}  }{B_4} \rho_n(r),  
 \end{eqnarray}
respectively, with $G_F$ is the Fermi coupling constant, $C_T$ is the T-PT coupling constant, 
{\boldmath$ {\vec \sigma}_n$}$= \langle \sigma_n \rangle \frac{{\bf \vec I}}{I}$ is the Pauli spinor of the nucleus, 
{\boldmath$ {\vec \gamma}_D$} represents for the Dirac matrices, $\rho_n(r)$ is the nuclear charge density, 
 $\mbox{\boldmath ${\vec S}$} = S \frac{{\vec{\bf I}}}{I}$ is the NSM and $B_4=\int_0^{\infty} dr r^4 \rho_n(r)$. 

\begin{figure}[t]
\includegraphics[width=8.5cm, height=5.5cm, clip=true]{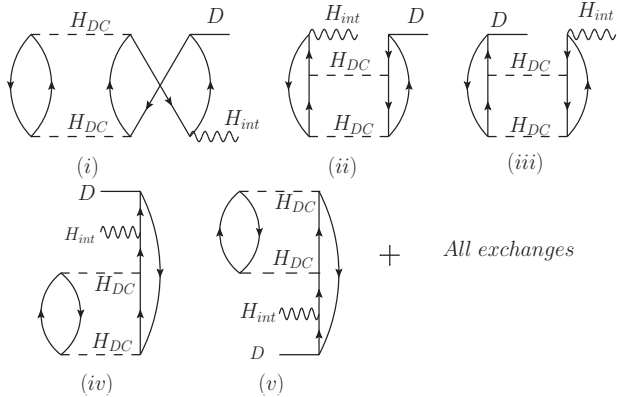}
\caption{Few important non-RPA diagrams from the MBPT method. Lines with up and down arrows are the occupied and 
unoccupied orbitals, respectively. $H_{int}$ corresponds to either $H_{e-N}^{T-PT}$, $H_{e-N}^{NSM}$ or $D$ 
in the evaluation of $d_A^{TPT}$, $d_A^{NSM}$ and $\alpha_d$ respectively.}
\label{fig1}
\end{figure}

The ground state wave function ($|\Psi_0^{(0)} \rangle$) of $^{225}$Ra is evaluated using the relativistic 
coupled-cluster (RCC) method considering the dominant atomic Dirac-Coulomb (DC) Hamiltonian ($H_{DC}$) by expressing 
$|\Psi_0 \rangle=e^{T^{(0)}}|\Phi_0 \rangle$ with the DF wave function $|\Phi_0 \rangle$ and the RCC operator $T^{(0)}$
that formulates excitation configurations acting upon $|\Phi_0 \rangle$ to account for the correlation effects. For
the determination of $d_A$ and $\alpha_d$, we also determine the first order wave functions ($|\Psi_0^{(1)} \rangle$s) 
with respect to $|\Psi_0^{(0)} \rangle$ by considering the respective P,T-odd interaction Hamiltonians and $D$ 
operator. In the RCC framework, we express \cite{yashpal-polz,yashpal-xe,bijaya-rn,yashpal-hg}
\begin{eqnarray}
 |\Psi_0^{(1)} \rangle = e^{T^{(0)}} T^{(1)} |\Phi_0 \rangle,
\end{eqnarray}
where $T^{(1)}$ is also the RCC operator similar to $T^{(0)}$ but generating all possible odd parity excitations from 
$|\Phi_0 \rangle$. For the calculations of both $|\Psi_0^{(0)} \rangle$ and $|\Psi_0^{(1)} \rangle$, we only allow 
all possible single and double excitation configurations (CCSD method). 
In this approach, both $d_A$ and $\alpha_d$ (both are symbolically denoted as $\mathcal{X}$) is determined by 
\cite{yashpal-xe,bijaya-rn,yashpal-hg}
\begin{eqnarray}  
\mathcal{X} &=& 2 \langle\Phi_0 |(\overline{D}^{(0)} T^{(1)})_{cc}|\Phi_0 \rangle              
\label{eq5}
\end{eqnarray}
where $cc$ stands for the terms those are closed and connected and $\overline{D}^{(0)}=e^{T^{\dagger{(0)}}}De^{T^{(0)}}$.
In our previous work \cite{yashpal-hg}, we have explained the procedure by which we take into account contributions 
from the non-truncative series $\overline{D}^{(0)}=e^{T^{\dagger{(0)}}}De^{T^{(0)}}$ by an iterative procedure. We present 
later results even with the intermediate truncation in this series as CCSD$^{(k)}$ method involving $k$ number of 
$T^{(0)}$ operators to highlight relevance of including higher order terms for accurate determination of $d_A$ and 
$\alpha_d$ (we shall refer to our CCSD$^{(\infty)}$ method to simply as CCSD method afterwards). In fact, our CCSD 
method includes contributions from lower order many-body perturbation theory (MBPT) and RPA. 
Therefore, the differences observed between the RPA and CCSD results are mainly because of the non-RPA contributions.
Few important lower order MBPT diagrams representing non-RPA contributions are shown in Fig. \ref{fig1}.

We optimize size of the basis to be able to carry out CCSD calculations within the available computational resources by 
performing RPA calculations (that requires less computational time and memory compared to the CCSD method) 
considering a different combination of low-lying single particle orbitals of $^{225}$Ra from the $s$, $p$, $d$ and $g$ 
angular momentum symmetries. The RPA results from few selective set of bases are given in Table \ref{tab1} and as seen,
taking orbitals from beyond basis set IV do not change the results in the present interest of accuracy. Our calculations 
from different methods using basis set IV are reported in Table \ref{tab2}. To estimate the uncertainties in our calculations, 
we estimate contributions from the orbitals of higher angular momentum symmetries that were not considered in the 
calculations, frequency independent Breit interaction and leading order quantum electrodynamics (QED) corrections 
(through the Uehling, Wichmann-Kroll and self-energy accounting potentials as described in \cite{flambaum-qed}) using
RPA. These contributions are also quoted separately in Table \ref{tab2} which show that the neglected contributions 
may not be larger than 1\%.

\begin{table}[t]
\caption{\label{tab1} Demonstration of convergence in the $\alpha_d$, $d_A^{TPT}$ and $d_A^{NSM}$ results, in units of 
$e a_0^3$, ($10^{-20}C_T \langle \sigma_n\rangle |e|cm$) and ($10^{-17}[S/|e|fm^3]|e|cm$) respectively with the finite 
bases, considering orbitals from the $s$, $p$, $d$, $f$ and $g$ angular momentum symmetries using RPA.} 
\begin{ruledtabular}
\begin{tabular}{lcccccccc}
\textrm{Basis} & $s$ & $p$ & $d$ & $f$ & $g$ & \multicolumn{3}{c}{\textrm{Results}} \\
  \cline{7-9}\\
               &  &  & & &  &   $\alpha_d$ &$d_A^{TPT}$ & $d_A^{NSM}$\\ 
               \hline
 I           &  10 & 9 & 8 & 5 & 4 & 295.49 & $-$11.60 & $-$5.58  \\
 II          &  12 & 11 & 10 & 7 & 6 &296.79& $-$14.81 & $-$7.20  \\
 III         &  14 & 13 & 12 & 9 & 7 &296.84& $-$16.10 & $-$7.84 \\
 {\bf IV}    &  {\bf 17} & {\bf 16} & {\bf 15} & {\bf 11} & {\bf 9} & {\bf 296.85} & {\bf $-$16.66} & {\bf $-$8.12} \\
 V           &  18 & 17 & 16 & 12 & 10 & 296.85& $-$16.66 & $-$8.12 \\
\end{tabular}
\end{ruledtabular}  
\end{table}

Though we have estimated uncertainties by taking into account various truncations in the methodologies of our numerical 
calculations, however it would at least be desirable to compare calculations of a physical quantity using the employed 
methods with its experimental result to test their capabilities for producing accurate result. As said before, 
evaluation of both $d_A$ and $\alpha_d$ have some similarities in the sense that their determination entails same 
angular momentum and parity selection criteria. On that account, calculations of $d_A$ and $\alpha_d$ are often 
performed together \cite{yashpal-xe,bijaya-rn,yashpal-hg, dzuba02,dzuba09,latha}, albeit it demands for different 
radial behavior of the wave functions for their estimation. Yet the experimental result of $\alpha_d$ in $^{225}$Ra 
to be realized, but many exclusive calculations using variants of RCC method report its value 
\cite{borschevsky,lim1,lim2}. Borschevsky {\it et al.} have used a CCSD method including important partial 
triples excitations perturbatively (CCSD(T) method) with the DC Hamiltonian and Gaunt term of the Breit interaction  
\cite{borschevsky}. The fundamental difference between their calculation with ours is that Borschevsky {\it et al.} 
use a finite field approach through a molecular code while we use the expectation value determination method describing 
wave functions in the spherical coordinate system. Consequently, Borschevsky {\it et al.} obtain large DF value 
compared to their CCSD(T) result. On the other hand, our CCSD result is larger than the DF value. 
Nevertheless, the final results in both the cases (that are suppose to converge towards the exact value) seem to be  
matching with the difference of about 2\%. On the other hand there are also two more calculations available using the methods similar 
to Borschevsky {\it et al.} but considering the scalar relativistic Douglas-Kroll (DK) Hamiltonian \cite{lim1} and 
small-core ten-valence electron scalar-relativistic pseudopotentials (ARPP) \cite{lim2}. Very good agreement between 
our CCSD result with the CCSD(T) values for $\alpha_d$ of $^{225}$Ra entails that our method, which is really an arduous 
procedure to implement in the spherical coordinate system, is adequate for the accurate determination of $d_A$ and 
$\alpha_d$. 

\begin{table}[t]
\caption{\label{tab2} Estimated $\alpha_d$, $d_A^{TPT}$ and $d_A^{NSM}$ (in same unit as of Table \ref{tab1}) results for 
$^{225}$Ra from various calculations. CCSD$^{(\infty)}$ values are considered as our final results and contributions 
from the Breit interaction ($\delta_B$), QED correction ($\delta_{QED}$) and truncated basis ($\delta_{basis}$) are 
accounted for estimating uncertainties to the CCSD results.}
\begin{ruledtabular}
\begin{tabular}{l|ccc|ccc}
Method of & \multicolumn{3}{c|}{\textrm{This work}}&\multicolumn{3}{c}{\textrm{Others}}\\
  \cline{2-4}  \cline{5-7}\\
 Evaluation  & \textrm{$\alpha_d$} & $d_A^{TPT}$& $d_A^{NSM}$ & $\alpha_d$ & $d_A^{TPT}$ &$d_A^{NSM}$ \\
\hline        \\
DF       & 204.13   &$-$3.46 &$-$1.85 &204.2 & $-$3.5$^a$ & $-$1.8$^a$  \\
         &          &        &        &200.6$^b$ &   &  \\
         &          &        &        &293.4$^c$       &  &   \\
         &          &        &        &300.57$^d$     &  &   \\
         &          &        &        &299.52$^e$     &  &   \\
RPA      &296.85    &$-$16.66&$-$8.12 &      & $-$17$^a$ & $-$8.3$^a$ \\
         &          &        &        &291.4$^b$  &$-$16.59$^b$  &  \\
         &          &        &        &297.0$^f$ &       &$-$8.5$^f$  \\
CI+MBPT  &          &        &        &      &$-$18$^a$  &$-$8.8$^a$  \\       
         &          &        &        &229.9$^f$      &  &  \\
& & \\
CCSD$^2$ &253.04    &$-$10.40&$-$6.94 & \\
CCSD$^4$ &242.02    &$-$9.49 &$-$6.52 & \\
CCSD$^{(\infty})$&247.76&$-$10.04&$-$6.79 &          &       &\\
CCSD(T) &          &        &        &242.8$^c$         &    &     \\  
   &          &        &        &248.56$^d$ &       &   \\
   &          &        &       & 251.12$^e$         &    &   \\
& & \\
$\delta_{B} $  &~ 0.19   &~ 0.06   &~ 0.06 &  \\ 
$\delta_{QED}$ &$-$0.43&$-$0.16&$-$0.07 &  \\ 
$\delta_{basis}$&$-$0.03&$-$0.08&$-$0.05 &  \\
\end{tabular}                    
\end{ruledtabular}
\begin{tabular}{l}
$^a$\cite{dzuba09} \\
$^b$\cite{latha} \\
$^c$\cite{borschevsky} Corrections from the Gaunt term incorporated.\\
$^d$\cite{lim1} DK operator is used in the calculations.\\
$^e$\cite{lim2} ARPP is considered in the calculations.\\
$^f$\cite{dzuba02} Corrections from the RPA included.\\
\end{tabular}
\end{table}  

After having a satisfactory comparison between the $\alpha_d$ values from the CCSD methods, we also analyze both the $d_A$ 
and $\alpha_d$ results from Table \ref{tab2} among other approximated methods using which $d_A$ of $^{225}$Ra were 
previously evaluated. Among the other previous theoretical studies, the combined configuration interaction and 
leading order many-body perturbation theory (CI+MBPT method) with some contributions through time-dependent DF method 
(equivalent to our RPA) employed by Dzuba and Coworkers seem to include more physical effects in the evaluation of 
both $d_A$ and $\alpha_d$ \cite{dzuba02,dzuba09}. In this approach, they have used a $V^{N-2}$ potential, $N=88$ is the
total number of electrons of $^{225}$Ra, to generate the single particle orbitals in contrast to our $V^N$ potential. 
Unlike our methods, the correlation effects from different electrons are not accounted in equal footing in these 
calculations. Also, Latha and Amjith have employed RPA with $V^N$ potential to report $d_A$ due to T-PT interaction and
$\alpha_d$ \cite{latha}. Results at RPA level match well between all these calculations. The CI+MBPT results of Dzuba 
and Coworkers, however, give larger $d_A$ but smaller $\alpha_d$ values compared to the RPA calculations. On the 
otherhand, correlation effects through CCSD method (especially the all order non-RPA contributions) reduce significantly
both $d_A$ and $\alpha_d$ with respect to the RPA results. It should be reminded that our CCSD method 
already includes RPA result. The $d_A^{TPT}$ and $d_A^{NSM}$ values from the CCSD method are reduced by 45\% and 23\%, 
respectively, compared to the CI+MBPT results for $^{225}$Ra even though difference between the $\alpha_d$ value from 
both the calculations are only about 10\%. We also present CCSD$^{(2)}$ and CCSD$^{(4)}$ results in Table \ref{tab2} to 
demonstrate the role of the correlation effects to improve the results gradually. 
%

Combining our final CCSD results with the measured $d_A$($^{225}$Ra) \cite{lu}, we get an upper bound on NSM as
$S \textless  7.4 \times 10^{-6} |e|\text{fm}^3$. Similarly with the knowledge of $\langle \sigma_n\rangle$
from the nuclear calculation, upper bound on $C_T$ can also be obtained. Two sophisticated nuclear structure 
calculations by considering the octupole deformed Wood-Saxon potential \cite{spevak-nuc} and the odd-A Skyrme mean 
field theory \cite{Dobaczewski} have been carried out to describe the P,T-odd interactions in the $^{225}$Ra nucleus. 
These calculations parameterize $S$ in terms of the pion-nucleon-nucleon ($\pi$NN) couplings. However, they provide 
different combinations of the $\pi$NN couplings to express $S$. Recently Engel {\it et al.} in their recent review give 
the best value for $S$ from these two calculations in terms of the $\pi$NN couplings as \cite{engel}
\begin{eqnarray}
 {\boldmath S}&=& 13.5[ -1.5 \bar{g}_0 + 6.0 \bar{g}_1 - 4.0 \bar{g}_2]\mbox{ $|e|$ fm}^3 ,
\end{eqnarray}
where $\bar{g}_{i=0,1,2}$ are the isospin components of the P,T-odd $\pi$NN coupling constants. Thus, combining the 
above result with our limit on $S$, we infer bounds on these couplings as $|\bar{g}_0|\textless 3.6\times 10^{-7}$
and $|\bar{g}_1|\textless 9.1\times 10^{-8}$. Further from the relations $|\bar{g}_0|=0.018(7) \theta_{QCD}$ 
\cite{dekens} and $|\bar{g}_1|=2\times 10^{-12} (\tilde{d}_u-\tilde{d}_d)$ \cite{pospelov1}, we extract the upper 
limits as $|\theta_{QCD}| \textless 2.0 \times 10^{-5}$ and  $|\tilde{d}_u - \tilde{d}_d| < 4.6 \times 10^{-22} |e| 
\text{cm}$. Though these bounds are not competitive at present against those obtained using the EDM study of 
$^{199}$Hg \cite{yashpal-hg}, however as said before rapid improvement in the EDM measurement of $^{225}$Ra is 
anticipated \cite{lu}. It, therefore, appears very likely that limits on the above quantities can become more 
stringent with the upcoming improved EDM measurement of $^{225}$Ra and when the nuclear calculations are advanced
further. These limits would definitely shed light on the new physics beyond the SM. 

We thank Professor B P Das for encouraging us to carry out this study. We are grateful to Professor Z -T Lu for 
informing us their $^{225}$Ra EDM experimental result. The computations were carried out using 3TFLOP HPC cluster 
of PRL, Ahmedabad.

\end{document}